\documentclass[aps,pre,superscriptaddress,twocolumn,longbibliography]{revtex4-2}

\usepackage{graphicx}
\usepackage{dcolumn}
\usepackage{bm}
\usepackage{epstopdf}
\usepackage{algorithm}
\usepackage{algpseudocode}
\usepackage[colorlinks]{hyperref}
\usepackage{amsmath,amsthm,amssymb}
\usepackage{float}
\usepackage{epsfig} 
\usepackage{mathrsfs}
\usepackage{multirow}
\usepackage[all]{xy}
\usepackage{pbox}
\usepackage{verbatim}
\usepackage{braket}
\usepackage{mathtools}
\usepackage{tikz}
\usepackage{xcolor}
\usepackage{xfrac}
\usepackage{cleveref}
\usepackage{todonotes}
\usepackage[utf8]{inputenc}
\usepackage{soul}

\newcommand{\singlefigure}{8.6cm}
\newcommand{\fullfigure}{17.8cm}
\newcommand{\params}{\bm{\theta}} 
\newcommand{\trajparams}{\bm{\Theta}} 

\begin{document}

\title{Minibatch training of neural network ensembles via trajectory sampling}

\author{Jamie F. Mair}
\email{Jamie.Mair@nottingham.ac.uk}
\affiliation{School of Physics and Astronomy, University of Nottingham, Nottingham, NG7 2RD, UK}
\author{Luke Causer}
\affiliation{School of Physics and Astronomy, University of Nottingham, Nottingham, NG7 2RD, UK}
\affiliation{Centre for the Mathematics and Theoretical Physics of Quantum Non-Equilibrium Systems,
University of Nottingham, Nottingham, NG7 2RD, UK}
\author{Juan P. Garrahan}
\affiliation{School of Physics and Astronomy, University of Nottingham, Nottingham, NG7 2RD, UK}
\affiliation{Centre for the Mathematics and Theoretical Physics of Quantum Non-Equilibrium Systems,
University of Nottingham, Nottingham, NG7 2RD, UK}

\date{\today}

\begin{abstract}

Most iterative neural network training methods use estimates of the loss function over small random subsets (or {\em minibatches}) of the data to update the parameters, which aid in decoupling the training time from the (often very large) size of the training datasets. Here, we show that a minibatch approach can also be used to train neural network ensembles (NNEs) via trajectory methods in a highly efficient manner.
We illustrate this approach by training NNEs to classify images in the MNIST datasets. This method gives an improvement to the training times, allowing it to scale as the ratio of the size of the dataset to that of the average minibatch size which, in the case of MNIST, gives a computational improvement typically of two orders of magnitude. We highlight the advantage of using longer trajectories to represent NNEs, both for improved accuracy in inference and reduced update cost in terms of the samples needed in minibatch updates. 
\end{abstract}

\maketitle

\section{Introduction}

Traditional machine learning (ML) applications aim to train a single model, usually by adjusting the parameters that define a complex function approximator like a neural network (NN), to perform well on some desired outcome as measured by a proxy loss function. A high performance on an appropriate loss function will entail a high performance on the metrics one cares about, for example accuracy in a classification problem \cite{Goodfellow-et-al-2016}. There is strong empirical evidence from numerical experiments that increasing the scale of single models improves performance, as for example in the timely class of large language models (LLMs) \cite{Rae-2022-Scaling}.

However, to counteract the seemingly ever-increasing size of LLMs, there has also been significant work towards devising smaller models with similar capabilities in order to reduce the computational cost of training and of inference. A notable recent example is Stanford's Alpaca \cite{alpaca}, based on LLaMA \cite{touvron2023llama,wang2022selfinstruct}, which can match GPT3.5 \cite{Liu2023} despite being over an order of magnitude smaller. Another possibility is to replace one large model by an {\em ensemble} of smaller models which can provide similar or better inferences while also being less costly to train and evaluate \cite{wen2020batchensemble,wang2022wisdom}. This is the class of problems we focus on here.

Recently, we introduced an approach to train collectively an ensemble of models \cite{Mair2022}, in particular neural network ensembles (NNEs) where predictions at inference time are aggregated in a committee-like fashion (for classification, the ensemble prediction is the most voted for option, while for scoring, the ensemble prediction is the mean ensemble score). In Ref.~\cite{Mair2022} we defined an NNE in terms of the trajectory of the model parameters under a simple (discrete in time, diffusive in parameter space) dynamics, and trained it by biasing the trajectory that defines the NNE towards a small time-integrated loss. That is, once training is converged, the NNE corresponds to a discrete trajectory of the model parameters sampled from a distribution of trajectories exponentially ``tilted'' to have low time-integrated loss. This approach is borrowed from the study of glassy systems \cite{chandler2010dynamics}, where biasing dynamics according to time-integrated observables (e.g. the dynamical activity \cite{garrahan2007dynamical,jack2011preparation}) is known to access low energy states for the configurations in the trajectory. Such low-loss trajectories can be accessed via importance sampling in trajectory space, such as transition path sampling (TPS) \cite{bolhuis2002transition} as adapted to stationary dynamics and large deviation problems \cite{hedges2009dynamic}. The ensuing trained NNE is a collection of NN models correlated by the underlying dynamics of the parameters and with a low value of the total loss due to the tilting. 

While Ref.~\cite{Mair2022} provides a proof of principle of the trajectory sampling approach, it suffers from a significant computational bottleneck: importance sampling is a Monte Carlo scheme on trajectories, where updates are determined according to changes in the (time-integrated) loss evaluated over the {\em whole training set}, so that each Monte Carlo iteration scales with the size of the training data. For example, when training for the textbook MNIST digit classification problem in Ref.~\cite{Mair2022}, we used only a small amount of the entire training dataset ($2048$ samples from the available $60000$) to make the problem tractable for a comprehensive study. On the contrary, it is well known that ML models generalise poorly with small datasets \cite{Goodfellow-et-al-2016}. This computational limitation makes the method of Ref.~\cite{Mair2022} impractical for more complex tasks. This has to be contrasted with gradient descent \cite{Goodfellow-et-al-2016}, where there is no need to sample faithfully from a distribution, so that the gradient of the loss can be estimated efficiently only on very small subsets of training data, known as {\em minibatches}, giving rise to stochastic gradient descent (where the noise from the difference between the minibatch estimate and the full loss  actually helps convergence to a good local minimum \cite{Goodfellow-et-al-2016}).

In this paper, we resolve the problem above by implementing a minibatch method in the trajectory sampling used in the training of the NNEs. We build on the approach of Ref.~\cite{Seita2018} for doing Monte Carlo sampling with small data batches. We show that our new method reduces the training cost by a factor given by the ratio of the average minibatch size (which we determine in an adaptive manner) to the size of the dataset. 
We illustrate this more efficient method on MNIST classification (using the whole MNIST dataset), showing in this case a computational gain of about two orders of magnitude. Our minibatch approach also allows us to highlight the key features of the trajectory NNE method, showing the advantage of using longer trajectories to represent NNEs both in terms of accuracy and data requirement for training. 

The rest of the paper is organised as follows. In Sec.~II we describe the theory, reviewing the idea of NNEs as trajectories of a stochastic dynamics, training as tilted trajectory sampling, and the central approach to perform mini-batch trajectory Monte Carlo. In Sec.~III we present the adaptive minibatch trajectory sampling method for training NNEs. We illustrate the method with two examples in Sec.~IV, an exactly solvable linear perceptron, and the full MNIST digit classification problem. In Sec.~V we give our conclusions, and further technical details are provided in the Appendices. 
\section{Theoretical background}

\begin{figure}
    \centering
    \includegraphics[width=\singlefigure]{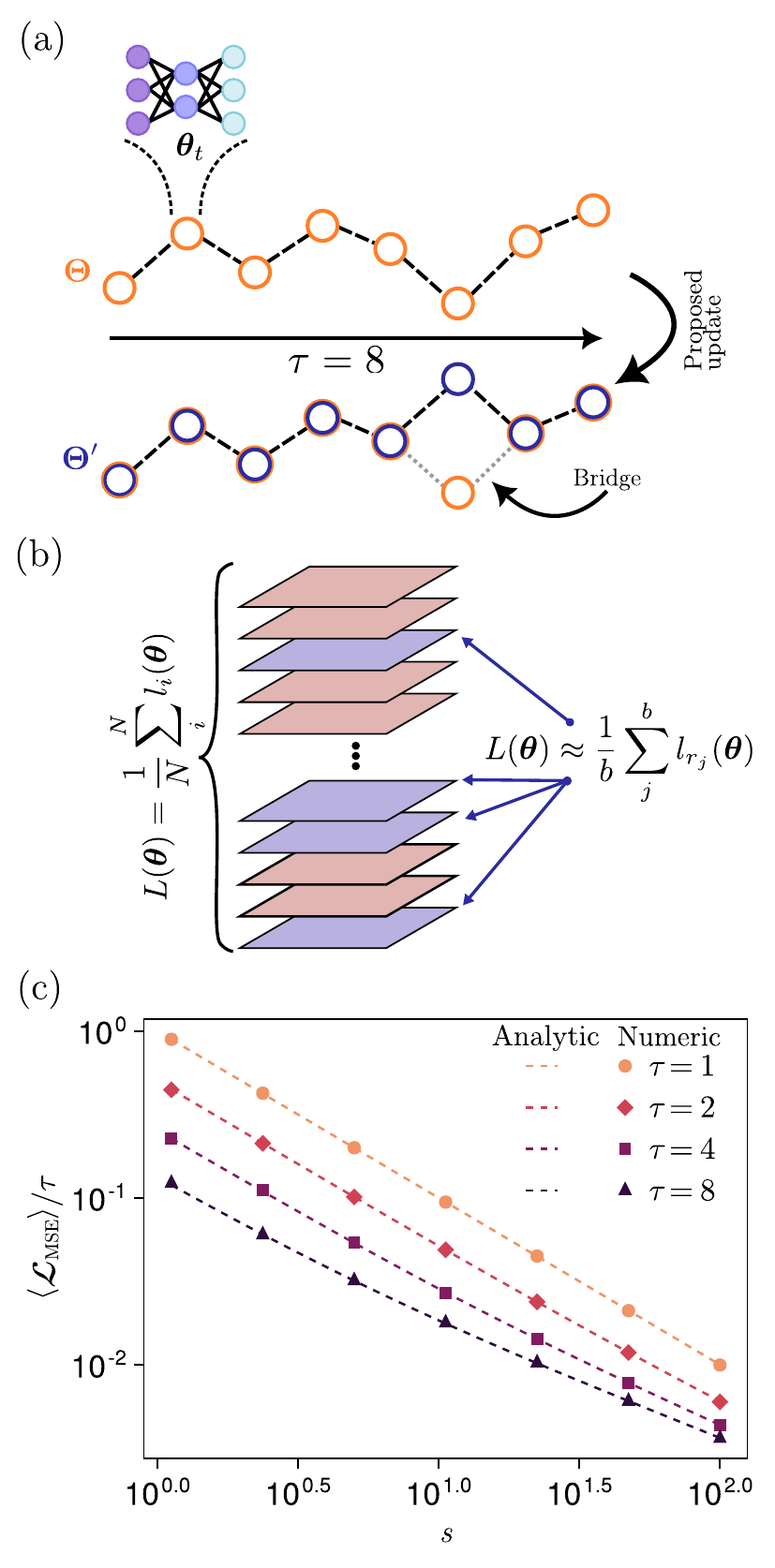}
    \caption{\label{fig:explanatory_combined_figure} 
    (a) An NNE as a stochastic trajectory, and a sketch of trajectory sampling. Each state in the trajectory corresponds to one NN model in the NNE. The proposed path sampling update from trajectory $\trajparams$ to trajectory $\trajparams'$ is via a stochastic bridge in which only one model is modified. 
    (b) The loss function of a model with parameters $\params$. Each layer represents a data element. The loss, $L(\params)$, is given by the average of the individual loss for each of the elements of the dataset, $\{ l_i(\theta) \}_{1:N}$. The minibatch estimate of the loss is instead the average over a random selection $\{ r_j \}_{1:b}$ of $b$ data points in the dataset.
    (c) Exactly solvable linear perceptron: mean NNE loss $\langle {\mathcal L} \rangle$
    per NNE size $\tau$, as a function of $s$ for various $\tau$. Lines are analytical results. Symbols are the numerical results using the minibatch TPS algorithm for training (with $20\times10^6$ TPS epochs as a ``burn-in'' followed by $20\times 10^6$ epochs for convergence of training).
    }
\end{figure}

\subsection{Neural network ensemble as a trajectory of neural networks}

In Ref.~\cite{Mair2022}, we proposed that a NNE could be obtained by evolving the parameters of a NN model under a suitable stochastic dynamics, where the NNE is composed of the sequence of NNs in time. If, at time step $t$, the NN is defined by $\params_t$, this dynamics would give rise to a trajectory $\params_1 \to \params_2 \to \ldots \to \params_\tau$, with the NNE as the set of visited models under the dynamics, $\trajparams=[\params_1, \params_2, \ldots, \params_\tau]$. As the aim is to minimise the loss over the ensemble
\begin{equation}
    {\mathcal L}(\trajparams) 
    = 
    \sum_{t=1}^\tau L(\params_t)
    \label{loss} 
\end{equation}
where $L(\params_t)$ is the standard loss for the $t$-th model (see below for a specific form of the loss), training is equivalent to finding a suitable dynamics whose typical trajectories are those with low time-aggregated loss, ${\mathcal L}(\trajparams)$.

Once trained, this dynamics is defined in terms of (in general time-dependent) stochastic dynamics ${\mathcal M}(\tau, \sigma, s) \equiv \{ M_{t; \sigma, s} \}_{t=1}^{\tau-1}$, where
$M_{t; \sigma, s}(\params' | \params)$ are the transition probabilities at each time step, 
such that the NNE corresponds to a trajectory generated using dynamics ${\mathcal M}(\tau, \sigma, s)$.
This approach is illustrated in Fig.~\ref{fig:explanatory_combined_figure}(a): the NNE is a discrete-time trajectory, where each state along the trajectory corresponds to one of the NNs that form the ensemble. Starting from the first model, $\params_1$, each subsequent model is sampled according to ${\mathcal M}(\tau, \sigma, s)$. Three hyperparameters determine the dynamics that produces the NNE: the final time $\tau$ sets the number of models in the ensemble; $\sigma$ sets the ``stiffness'' of the chain (see below for details), that is, how correlated subsequent models are to each other, with small $\sigma$ corresponding to large stiffness; and $s$ controls the level of the overall loss, with larger $s$ corresponding to lower loss. The central idea is that a correlated chain of models generated as a trajectory from dynamics ${\mathcal M}(\tau, \sigma, s)$ at large $s$ provides a well trained NNE \cite{Mair2022}.

\subsection{Learning as a trajectory sampling problem}

Obtaining a suitable dynamics ${\mathcal M}(\tau, \sigma, s)$ that produces well-trained NNEs as its typical trajectories is a difficult task. We can however resolve this problem by means of trajectory sampling techniques \cite{Mair2022}. Consider as a starting point an untrained dynamics ${\mathcal M}(\tau, \sigma, 0)$ 
with the same transition probabilities 
at every time step $t$ \cite{Mair2022} 
\begin{align}
    \label{eq:M0}
    M_\sigma(\params_{t+1} | \params_{t}) 
        \propto
            \exp 
            \left [ 
                -\frac{1}{2 \sigma^2}(\params_{t}-\params_{t+1})^2 
            \right ] ,
\end{align}
with $\int_{\params'} M_\sigma(\params' | \params) = 1$. 
This dynamics corresponds to a discrete-time Gaussian diffusion process that knows nothing about the loss \eqref{loss}. As such, a typical trajectory drawn from it will correspond to a random (and therefore untrained) NNE. 
As indicated above, the parameter $\sigma$ sets the variance of the diffusive steps, so that for smaller $\sigma$ subsequent models are more correlated, while for $\sigma \to \infty$ all the models of the chain are uncoupled.

The dynamics \eqref{eq:M0} produces an {\em ensemble of trajectories} (and therefore an ensemble of NNEs) with each trajectory having probability 
\begin{align}
    \label{eq:unbiased1}
    P(\trajparams; \sigma) 
        = 
        \frac{1}
        {\mathcal{Z}_{\tau}(\sigma)}
        p(\theta_1)
        \prod_{t=1}^{\tau-1} 
        \exp 
        \left [ 
            -\frac{1}{2 \sigma^2}(\params_{t}-\params_{t+1})^2 
            \right ],
\end{align}
given by the product of the $M_\sigma$ at each step. 
Here $p(\theta_1)$ is the probability used to draw the first model, and ${\mathcal{Z}_{\tau}(\sigma)}$ a normalisation constant (the ``partition sum'' of the trajectory ensemble). In order to obtain trajectories with low overall loss, what we aim is to define a new trajectory ensemble that is exponentially ``tilted'' with respect to \eqref{eq:unbiased1}, 
as is standard in large deviation studies of dynamics (e.g., Ref.~\cite{Touchette2009}), that is \cite{Mair2022}
\begin{align}
    \label{eq:tilted-distribution}
    P(\trajparams; \sigma, s) 
        = 
        \frac{1}
        {\mathcal{Z}_{\tau}(\sigma,s)}
        e^{-s {\mathcal L}(\trajparams)}
        P(\trajparams; \sigma) . 
\end{align}
For large $s$, a typical trajectory from this ensemble will correspond to a NNE with low overall loss. The learned dynamics ${\mathcal M}(\tau, \sigma, s)$ of the previous subsection would be the dynamics that produces trajectories distributed according to the tilted distribution \eqref{eq:tilted-distribution}. 

One way to avoid having to determine the ${\mathcal M}(\tau, \sigma, s)$ dynamics explicitly is to directly sample trajectories from the tilted distribution \eqref{eq:tilted-distribution}. In this way, convergence of the training, that is, finding ${\mathcal M}(\tau, \sigma, s)$, coincides with convergence of the trajectory sampling of \eqref{eq:tilted-distribution}, as we do here by means of an importance sampling method in trajectory space based on transition path sampling (TPS) \cite{bolhuis2002transition}.

\subsection{Monte Carlo in trajectory space}

Consider a Monte Carlo scheme for sampling trajectories, specifically a Metropolis-Hastings approach \cite{bolhuis2002transition}: given a current trajectory $\trajparams$, the probability to change to a new trajectory $\trajparams'$ is given by 
\begin{equation}
    p(\trajparams'|\trajparams) = g(\trajparams'|\trajparams) A(\trajparams', \trajparams),
\end{equation}
where the factor $g(\trajparams'|\trajparams)$ 
is the probability to propose the move, and $A(\trajparams', \trajparams)$ is that to accept it. 
For the above to converge to \eqref{eq:tilted-distribution} we need to impose that it obeys detailed balance with respect to \eqref{eq:tilted-distribution}, which implies 
\begin{equation}
    \label{eq:full-detailed-balance-condition}
    \frac{
        A(\trajparams', \trajparams)
        }
        {
        A(\trajparams, \trajparams')
        } 
    = 
    \frac{
        P(\trajparams'; \sigma) 
        g(\trajparams|\trajparams')
        }
        {
        P(\trajparams; \sigma) 
        g(\trajparams'|\trajparams)
        } 
\end{equation}
If the proposed moves obey detailed balance with respect to the original untilted dynamics \eqref{eq:unbiased1}, 
\begin{equation}
    \label{eq:propratio}
    \frac{
            g(\trajparams'|\trajparams)
        }
        {
            g(\trajparams|\trajparams')
        } 
    =  
    \frac{
            P(\trajparams'; \sigma)
        }
        {
            P(\trajparams; \sigma)
        } , 
\end{equation}
then the acceptance ratio reduces to 
\begin{equation}
    \label{eq:mh-acceptance}
    \frac{
        A(\trajparams', \trajparams)
        }
        {
        A(\trajparams, \trajparams')
        } 
    = 
    e^{
            -s 
            \left[
                {\mathcal L}(\trajparams')
                -
                {\mathcal L}(\trajparams)
                \right]
    }
\end{equation}

In standard TPS, \eqref{eq:propratio} is realised by proposing trajectories by simply running the original dynamics \eqref{eq:unbiased1} (via ``shooting'' or ``shifting'' moves, see Ref.~\cite{bolhuis2002transition}). This approach, however, carries an exponential cost in the time extent of trajectories, since the loss difference in the exponent of \eqref{eq:mh-acceptance} scales linearly with time. 
This can be mitigated \cite{Mair2022} by proposing small changes in a trajectory, see Fig.~\ref{fig:explanatory_combined_figure}(a): the proposed trajectory is one where only the state at one time is modified; as this has to obey \eqref{eq:propratio}, it has to be done as a {\em Brownian bridge} \cite{Grela2021,De-Bruyne2021}. That is, as conditioned dynamics starting in the previous state and retuning to the state after the one changed (see Ref.~\cite{Mair2022} for details).

\section{Minibatch path sampling}

While the Brownian bridge version of TPS ameliorates the exponential-in-time cost in the trajectory sampling, there is another source of computational slowness coming from the evaluation of the trajectory loss, cf.\ \eqref{eq:mh-acceptance}. 
With the Brownian bridge TPS, Fig.~\ref{fig:explanatory_combined_figure}(a), only a single model changes between the current and proposed trajectory, and the change in trajectory loss in \eqref{eq:mh-acceptance} is therefore given by that the change of that model's loss. 
If this change is at time $t$, this requires the evaluation of $L({\params'_{t}})$, which at training is the average of the loss under that model for each of the $N$ training data points,
\begin{equation}
    L({\params'_{t}}) 
        = 
        \frac{1}{N}
        \sum_{i=1}^N
            l_i(\params'_t)
\end{equation}
where $l_i(\params_t)$ is the loss for the inference for data point $i$. This means that in each Monte Carlo iteration, computing the change in loss inevitably scales with the training set size $N$ (together with a cost that depends on the size and architecture of the NN being considered). 
This evaluation can therefore become computationally infeasible for larger datasets. For example, in Ref.~\cite{Mair2022}, we had to reduce the training dataset by almost an order of magnitude to show a proof-of-principle of the method with for the MNIST classification problem. 

In contrast to gradient descent, one cannot simply replace the loss over the whole training set for an estimate based on a small subset, or minibatch. For gradient descent, the error that this introduces becomes a source of noise, converting it into stochastic gradient descent (and its adaptive variants \cite{Goodfellow-et-al-2016,Lecun1998,Bottou2010}). This in turn gives rise to the usual advantages that an exploit/explore strategy brings, in this case to minimise the loss locally descending the gradient versus exploration of the loss landscape. 
Since Monte Carlo aims to sample from a distribution, Eq.~\eqref{eq:tilted-distribution} in our case, 
a straightforward replacement of the loss by a minibatch approximation would lead to failure of the necessary detailed balance condition.

This problem has been considered before in the context of Bayesian inference, where so-called ``tall datasets'' make Monte Carlo inefficient, see e.g., Refs.~\cite{korattikara2014austerity,bardenet2014towards,bardenet2017on-markov}. In what follows we build on the approach put forward in Ref.~\cite{Seita2018} to develop an adaptive minibatch trajectory sampling method.

\subsection{Monte Carlo with minibatches}

We first describe the scheme of Ref.~\cite{Seita2018} in the context of the Monte 
Carlo annealing of a system with degrees of freedom $\trajparams$ (a NNE in our case) and target distribution \eqref{eq:tilted-distribution}, and in the next subsection we extend the approach to integrate it with TPS in an adaptive manner.

Let us define the quantity $\Delta(\params', \params)$ as the logarithm of the change in weight under a proposed move,
\begin{equation}
    \label{eq:delta}
    \Delta(\trajparams', \trajparams) 
    = 
    -s 
        \left[
                {\mathcal L}(\trajparams')
                -
                {\mathcal L}(\trajparams)
                \right],
\end{equation}
and choose our acceptance function as
\begin{equation}
    \label{eq:logistic-acceptance-fn}
    A(\trajparams', \trajparams)=(1+e^{\Delta(\trajparams', \trajparams)})^{-1},
\end{equation}
which satisfies the detailed balance condition \eqref{eq:mh-acceptance}. Monte Carlo works by generating a proposed move from $g(\trajparams'|\trajparams)$, and then accepting the move if 
\begin{equation}
    \label{eq:original-acceptance-test}
    A(\trajparams', \trajparams) > V,
\end{equation}
where $V$ is a uniformly distributed random number, $V \sim \mathcal{U}(0, 1)$. As \eqref{eq:logistic-acceptance-fn} is a logistic (or sigmoid) function whose inverse is also its derivative, we can equivalently write the acceptance test as $\Delta(\trajparams', \trajparams) > X_\text{log}$, where $X_\text{log}$ is a logistically sampled random variable. As this distribution is symmetric around zero, we can equally write the test \eqref{eq:original-acceptance-test} as 
\begin{equation}
    \label{eq:test}
    \Delta(\trajparams', \trajparams) + X_\text{log} > 0.
\end{equation}
The loss ${\mathcal L}(\trajparams)$ that enters in \eqref{eq:delta} is the average of the loss over the entire training data set
\begin{equation}
    \label{lossall}
    {\mathcal L}(\trajparams) 
    = 
    \frac{1}{N}
    \sum_{i=1}^N
        {\mathcal L}_i(\trajparams) 
    = 
    \frac{1}{N}
    \sum_{i=1}^N
    \sum_{t=1}^\tau
        l_i(\params_t) 
    ,
\end{equation}
where ${\mathcal L}_i(\trajparams)$ is the (trajectory) loss for the inference on data point $i$. Consider now an approximation of the loss difference in terms of a random minibatch of size $b$ 
\begin{equation}
    {\mathcal L}(\trajparams) 
    \approx
    \frac{1}{b}
    \sum_{j=1}^b
        {\mathcal L}_{r(j)}(\trajparams) , 
\end{equation}
where $r(j)$ specifies a random permutation of the indices of the elements in the training dataset, cf.\ Fig.~\ref{fig:explanatory_combined_figure}(b). In terms of the above we can define an approximation to \eqref{eq:delta} \cite{Seita2018},
\begin{equation}
    \label{eq:deltastar}
    \Delta^*(\trajparams, \trajparams') = - \frac{s}{b} 
    \sum_{j=1}^{b} 
    \left[
    {\mathcal L}_{r(j)}(\trajparams')
    -
    {\mathcal L}_{r(j)}(\trajparams) 
    \right] .
\end{equation}
If one could replace $\Delta(\trajparams, \trajparams')$ by $\Delta^*(\trajparams, \trajparams')$, then there would be a computational gain that would scale as the ratio of the size $N$ of the training dataset to that of the minibatch $b$ (as it is necessary to compute $\Delta(\trajparams, \trajparams')$ for each Monte Carlo iteration). The way to do so is as follows \cite{Seita2018}. 

Since the elements of the minibatch are chosen in an identical and independent manner, from the central limit theorem and for large enough $b$, we expect $\Delta^*$ to be normally distributed around $\Delta$ with some variance $\rho^2{(\Delta^*)}$. That is,  $\Delta^* = \Delta + X_{\text{norm}}$, where $X_{\text{norm}}$ is an approximately normal zero-mean random correction of variance $\rho^2{(\Delta^*)}$. As a logistically distributed random variable, $X_\text{log}$, is almost normally distributed, we can write $X_\text{log}$ as $X_\text{log} = X_\text{norm}+X_\text{corr}$, where $X_\text{corr}$ is the (hopefully small) correction to normality. Inserting this decomposition of $X_\text{log}$
into the acceptance test \eqref{eq:test}, we can replace $\Delta$ with the minibatch estimate:
\begin{equation}
    \label{eq:teststar}
    \Delta^*(\trajparams', \trajparams) + X_\text{corr} > 0.
\end{equation}
This new acceptance only depends on the minibatch estimates of the loss and --- if accurate --- will be efficient for $b \ll N$. The test \eqref{eq:teststar} will asymptotically give the correct acceptance distribution provided that (i) that the fluctuations $X_{\text{norm}}$ of $\Delta^*$ around $\Delta$ are normally distributed (which can be checked to adjust the size of $b$), and (ii) that the distribution $C_{\rm corr}(X; \rho)$ for the random correction $X_{\text{corr}}$ can be numerically computed with low error (see \cite{Seita2018} for analysis of the errors).

Thanks to the CLT, condition (i) is relatively easy to satisfy for a large enough minibatch sample size. If the batch size grows beyond the size of the dataset, we do not need this approximation and can use \eqref{eq:original-acceptance-test}. Condition (ii) holds when the sample error on $\Delta^*$ is sufficiently small \cite{Seita2018}.
In practice one can only compute the distribution of $C_{\rm corr}(X; \rho)$
accurately enough only for standard deviations of $X_{\text{norm}}$ 
such that $\rho \lesssim 1.1$ (see below for our implementation). 

As computing $C_{\rm corr}(X; \rho)$ is numerically expensive (see \Cref{app:correction-distribution} for details) for each empirical $\rho \approx 1$, we can instead compute a $\rho^2=1$ correction and then add a further ``normal correction'' with small variance $1-\rho^2$ to make a total normal random variable with fixed variance $1$. Putting all of these together, we get the minibatch acceptance test that we use
\begin{equation}
    \label{eq:minibatch-acceptance-test}
    \Delta^*(\trajparams, \trajparams') + X_\text{nc} + X_{\text{corr}} > 0 ,
\end{equation}
where $X_\text{nc}$ stands for the normal correction random variable, of zero mean and variance $1-\rho^2$.

\subsection*{Generalisation to enable training with TPS}

The TPS scheme that we use relies on proposing trajectory updates, cf.\ Fig.~\ref{fig:explanatory_combined_figure}(a), consisting of 
\textit{bridging} moves (for changes in the middle of the trajectory) and \textit{shooting} moves to get changes in the endpoints, see Ref.~\cite{Mair2022} for details. In either case, only a single model (i.e.\ a single time step) is altered, thus reducing the size of the update and improving the acceptance rate.

Once a candidate trajectory is proposed, the minibatch acceptance criterion of the previous subsection is applied. The specific steps are as follows, defining an adaptive minibatch scheme: 
\begin{itemize}
    \item
    We draw $m$ random samples from the training set, which are used to calculate an estimate of $\Delta^*(\trajparams, \trajparams')$ and $\rho^2(\Delta^*)$, cf.~\eqref{eq:deltastar}. If $\rho^2>1$, $m$ more samples are drawn without replacement, updating $\Delta^*$ and $\rho$ accordingly (and terminating if all samples are used). In this way, we form a minibatch of overall size $b$ such that the sample variance of $\Delta^*$ is strictly less than or equal to $1$.

    \item 
    We draw the random correction $X_\text{nc}$ from a normal distribution and $X_{\text{corr}}$ from $C_{\rm corr}(X; \rho)$. With these we use \eqref{eq:minibatch-acceptance-test} to accept or reject the proposed change to the trajectory. [If the total minibatch size $b$ equals $N$, then we use the original test \eqref{eq:original-acceptance-test} as $\Delta^*$ coincides with $\Delta$ in that case.]
    Unlike Ref.~\cite{Seita2018}, exact sampling of \eqref{eq:tilted-distribution} is not required to effectively train our NNE, and we do not test the normality assumption of $\Delta^*(\trajparams, \trajparams')$. This is equivalent to setting their threshold $\delta \rightarrow \infty$ in Ref.~\cite{Seita2018}. Justification for this simplifying choice is given in \Cref{app:normality-investigation}.
\end{itemize}

\begin{algorithm}[H]
	\caption{\label{alg:minibatch-training} Minibatch TPS Training}
	\begin{algorithmic}[1]
		\State \textbf{input} Initial trajectory $\trajparams_1$, dataset $\{x_1, \ldots, x_N \}$, trajectory length $\tau$, trajectory coupling $\sigma$, minibatch chunk size $m$, pre-computed correction $C_{\rho=1}(X)$ distribution, cut-off hyperparameters $c_0$ and $c_1$ and training epochs $E$.
		\State \textbf{output} Sequence of trajectories $\{\trajparams_1, \trajparams_1, \ldots, \trajparams_{E+1} \}$
        \For {$k \in [1,2,\ldots, E]$}
            \State Sample proposal $\trajparams'$ using $g(\trajparams'| \trajparams_{k}, \tau, \sigma)$ (i.e. shooting or bridging)
            \State Sample $\Delta^*(\trajparams', \trajparams_{k})$ and $\rho^2$ using $m$ randomly selected samples, without replacement
            \State $b \gets N$
            \While {$\rho^2>1$ \textbf{and} $b<N$, \textbf{and} $\max (\frac{|\Delta^*(\trajparams, \trajparams')|}{\rho} - c_1, 0) <= c_0$ }
                \State Select $m$ more randomly selected samples, without replacement and update estimates for $\Delta^*(\trajparams', \trajparams_{k})$ and $\rho^2$
                \State $b \gets b + m$
            \EndWhile

            \If {$b=N$}
                \State Sample random number $V\sim \mathcal{U}(0,1)$
                \If {$V<g(\Delta^*(\trajparams', \trajparams_{k}))$}
                    \State Accept with $\trajparams_{k+1} \gets \trajparams'$
                \Else
                    \State Reject with $\trajparams_{k+1} \gets \trajparams_{k}$
                \EndIf
                \State \texttt{\textbf{continue}}
            \EndIf

            \State Sample $X_\text{nc}\sim \mathcal{N}(0, 1-\rho^2)$ and $X_\text{corr}\sim C_{\rho=1}(X)$
            \If {$\Delta^*(\trajparams', \trajparams_{k}) + X_\text{nc} + X_\text{corr} > 0$}
                \State Accept with $\trajparams_{k+1} \gets \trajparams'$
            \Else
                \State Reject with $\trajparams_{k+1} \gets \trajparams_{k}$
            \EndIf
        \EndFor
	\end{algorithmic}
\end{algorithm}

One issue with this simple algorithm is that the minibatch size grows if the sample variance is much larger than $1$. The biasing parameter, $s$, scales the sample variance with $s^2$, while taking $b$ samples only reduces this variance by a factor of $b$. For fixed $\trajparams$ and $\trajparams'$, we would expect the minibatch size to change with $s$ to compensate for the increased sample variance. For $s \rightarrow \infty$, this would revert the minibatch method to the original full-dataset acceptance test. To avoid this, we introduce a \textit{cut-off test} which halts increasing the minibatch size when $\Delta^*$ is sufficiently far away from the origin and performs an alternative acceptance test. The alternative acceptance test is broken into two stages. Firstly, we approximate the acceptance function to be equal to $0$ when $x<-c_0$ and $1$ when $x>c_0$ for some positive constant $c_0$. We choose $c_0$ to be sufficiently high that the approximate acceptance function is unchanged for $-c_0\leq x \leq c_0$ without having to renormalise. Secondly, we choose another threshold for which the true mean $\Delta$ is approximately guaranteed to be within $\Delta^* \pm c_1 \rho$. We use $
    \max (\rho^{-1} | \Delta^*(\trajparams, \trajparams')| - c_1, 0) > c_0
$
as our cut-off acceptance test. In our experiments, using $c_1=10$ and $c_0=5$ gave good results. The final combined algorithm is given in \Cref{alg:minibatch-training}. Setting $c_0$ or $c_1$ to be high will make this cut-off less likely to be triggered, increasing minibatch size and therefore computational cost, however, setting them too close to $0$ will result in an inaccurate acceptance test which does not obey detailed balance.
\section{Examples of training NN ensembles via minibatch trajectory sampling}

We now apply the method of Sec.~III for the training of NNEs in two illustrative problems. The first is that of a linear perceptron, which is simple enough to be solved exactly, allowing us to directly compare our method with the expected results. The second is the more complex, but now standard, problem of MNIST digit classification \cite{deng2012mnist}.

\subsection{NNE of linear perceptrons}

We first test the method with a linear classification problem, also considered in Ref.~\cite{Mair2022}. This problem can be defined as follows: we generate a set of independent random $D$ dimensional points, $\{ \bm{x}_i \}_{i=1}^N$ (setting $x_N=1$), together with a $D$-dimensional random weight vector $\bm{w}$ that we use to assign labels $y_i = \bm{w} \cdot \bm{x}_i$ to each of the points $\bm{x}_i$. The aim is to train the parameters $\trajparams$ of an ensemble of $\tau$ linear perceptrons, where the prediction of the $t$-th perceptron for the $i$-th data point is $\params_t \cdot \bm{x}_i$. For training, we consider the mean-squared sample loss for each model in the NNE, which for data point $i$ reads
\begin{equation}
    l_i^\text{(MSE)}(\params_t) = 
    \frac{1}{2} 
    (y_i - \params_t \cdot \bm{x}_i)^2 .
\end{equation}
The NNE loss over the training dataset, cf.~\eqref{lossall}, in turn reads
\begin{equation}
    {\mathcal L}(\trajparams) 
    = 
    \frac{1}{N}
    \sum_{i=1}^N
    \sum_{t=1}^\tau
        \frac{1}{2} 
        (y_i - \params_t \cdot \bm{x}_i)^2
\end{equation}
We now implement the adaptive minibatch estimation of the trajectory loss described in Alg.~\ref{alg:minibatch-training} to train this NNE for various $\tau$ and $s$. Figure~\ref{fig:explanatory_combined_figure}(c) demonstrates that the numerics obtained in this way coincide with the analytic results from the exact trajectory distribution \eqref{eq:tilted-distribution} \cite{Mair2022}. This is an elementary proof-of-principle of the method. 

Exact distributions for $\bm{w}$ and $\bm{x}$, along with experimental hyperparameters are provided in \Cref{app:linear-perceptron}.

\begin{figure*}
    \centering
    \includegraphics[width=\fullfigure]{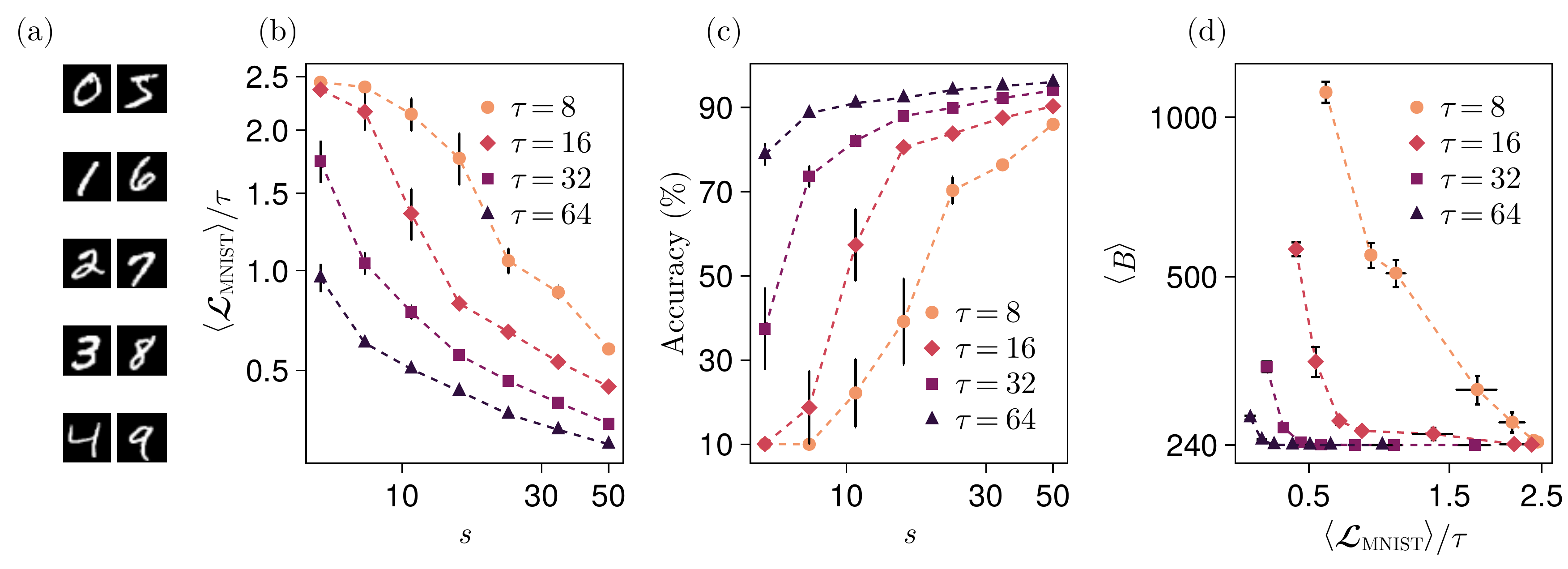}
    \caption{\label{fig:mnist_combined_figure} (a) Representative examples of the MNIST digit images used for training. 
    (b) Mean value of the time-averaged loss vs $s$. Each data point is calculated after $2 \times 10^{7}$ TPS epochs over the following $2 \times 10^{7}$ TPS epochs.
    (c) Final accuracy on the standard $10,000$ test images of a single trained NNE (via majority vote), taken at the end of the $4\times 10^7$ epochs.
    (d) Average batch size per epoch, $\langle b \rangle$ as a function of the converged mean NNE loss after $2\times10^7$ TPS epochs. The mean batch size per epoch is obtained from the $4\times10^7$ TPS training epochs.\newline All results and error bars are from averaging over $6$ independent runs with different initial trajectories.}
\end{figure*}

\subsection{NNE for MNIST digit classification}

The second problem we consider is that of an ensemble of models for classification of digits using the standard set of handwritten MNIST images \cite{deng2012mnist}, see Fig.~\ref{fig:mnist_combined_figure}(a). In this case, each NN in the NNE is a small convolutional neural network (CNN) whose architecture is described in \Cref{app:mnist}. For a given image $X$, one of these CNNs with parameters $\params$ provides the probability $y(k | X; \params)$ that the image corresponds to digit $k$, for $k = 0, \ldots, 9$. The appropriate loss function is the mean cross entropy, which for the data point $i$ and the $t$-th model reads
\begin{equation}
    l_i^\text{(MNIST)}(\params_t) = 
    - \sum_{k=0}^{9} 
    \delta_{z_i,k} \log {y(k|X_i; \params_t)},
\end{equation}
where $z_i$ is the true classification of $X_i$. The training loss for the NNE then reads
\begin{equation}
    {\mathcal L}(\trajparams) 
    = 
    - \frac{1}{N}
    \sum_{i=1}^N
    \sum_{t=1}^\tau
    \sum_{k=0}^{9} 
        \delta_{z_i,k} \log {y(k|X_i; \params_t)},
\end{equation}

\begin{figure}
    \centering
    \includegraphics[width=\singlefigure]{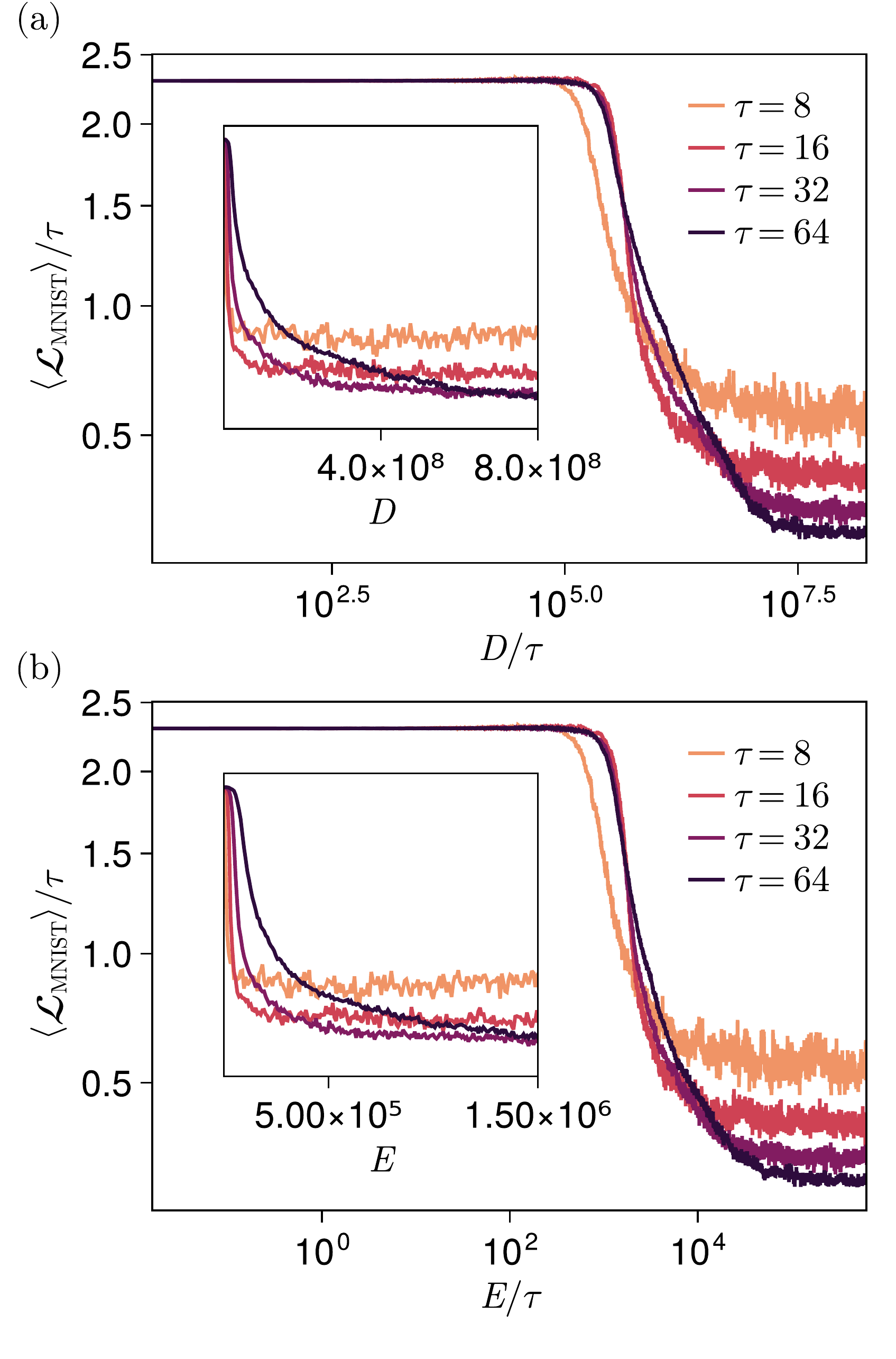}
    \caption{\label{fig:combined_loss_curves} 
    Training curves for MNIST classification NNE.
    (a) Average ensemble loss per model in the NNE as a function of cumulative data usage $D$, for $s=50$ and various NNE sizes $\tau$. Note that the abscissa is scaled by $\tau$, so curves are in terms of ``per-model'' epochs. In terms of $D/\tau$, for equivalent training time lower losses are reached for larger NNEs. (Loss curves has been down sampled for clarity.) Inset: same in linear $D$ scale. 
    (b) Same training curves but now plotted in terms of epochs $E$.
    }
\end{figure}

We train trajectories for a fixed number of epochs $E$ (i.e., TPS iterations), which we choose to be large enough for the trajectory loss to appear to converge, which we track in terms of the minibatch loss estimate to avoid further computational cost. One can observe convergence for a small number of sample runs in Figs.~\ref{fig:combined_loss_curves}(a, b). Specifically, we allow for a ``burn-in'' of the initial $20\times 10^6$ epochs, and subsequently observe the average trajectory loss for the next $20\times 10^6$ epochs. For each set of hyperparameters $(\tau, s)$ we perform six independent trainings starting each training run from a random initial seed trajectory. The time-averaged loss thus obtained is 
shown in Fig.~\ref{fig:mnist_combined_figure}(b) as a function of $s$ for various NNE sizes $\tau$. We note the following: (i) for every $\tau$ the  loss per model in the trained NNE decreases with $s$, as should be the case when converging to \eqref{eq:tilted-distribution}; (ii) the larger $\tau$ the lower the loss, indicating that the longer trajectories give rise to more accurate NNEs; (iii) there appears to be a transition from high to low loss with $s$ which could indicate (dynamical) phase coexistence, as seen in many other trajectory ensemble problems \cite{FOLENA2022128152}.

A similar trend to that of the loss is observed in the accuracy on the generalisation test set. In Fig.~\ref{fig:mnist_combined_figure}(c) we show the accuracy of the final ensembles obtained after all training epochs. We use the NNE to collectively make a prediction on each sample by letting each model ``vote'' for their predicted class, with the class with the most votes getting selected (in the event of a tie, the smaller digit is selected). We plot this ensemble accuracy, averaged over six independent runs, for the same hyperparameters of panel (b).

As a proxy for the computational cost of training, in Fig.~\ref{fig:mnist_combined_figure}(d) we show the average batch size per epoch $\langle b \rangle$ necessary for training to a certain value of the loss per model. Since the size of the training set is $N = 6 \times 10^4$, the ratio $N / \langle b \rangle$ gives the computational gain of using the minibatch method. From Figs.~\ref{fig:mnist_combined_figure}(b, c), we know that longer trajectories can yield a lower loss at smaller values of $s$. From Fig.~\ref{fig:mnist_combined_figure}(d), we see that longer trajectories (larger NNEs) are also computationally more efficient to train, requiring a smaller mean minibatch size, $\langle b \rangle$, than smaller NNEs for the same level of overall loss, showing a computational gain in excess of two orders of magnitude of the minibatch method to the original full loss method of Ref.~\cite{Mair2022}.

\section{Conclusions}

In this paper we have presented a variant of the minibatch Monte Carlo method of Ref.~\cite{Seita2018} adapted to the sampling of trajectories that correspond to neural network ensembles \cite{Mair2022}. We have shown that this technique can be used to train NNEs via trajectory sampling to give an improvement in computational efficiency up to two orders of magnitude. While we have focused, for concreteness, on supervised learning applications, we note that an adaptive trajectory sampling technique like the one presented here should be also very useful in Monte Carlo based {\em reinforcement learning} (RL), where datasets do not have a fixed size. We expect that this method will provide a stable training technique on these RL problems, which have exhibited brittle behaviour when continuously trained on changing objectives \cite{sutton_barto2018reinforcementlearning, bengio2020interference, MCCLOSKEY1989109, khetarpal2022towards}. 

Our results here add to the growing number of recent works studying the training dynamics of NNs from the statistical mechanics point of view, see e.g., Refs.~\cite{mei2018mean,rotskoff2022trainability,Whitelam2021,veiga2022phase,adhikari2023machine}. Most of these consider the training of a single NN in terms of a stochastic dynamics akin to thermal annealing, cf.\ Ref.~\cite{Whitelam2021}. In contrast, our approach based on sampling trajectories of NNs shares more similarities to training by quantum annealing, see for example Refs.~\cite{baldassi2018efficiency,lami2023quantum}. Note that this similarity is not referring to actual unitary dynamics, but to the fact the computation of a trajectory ensemble in \eqref{eq:tilted-distribution} is similar to that of a quantum partition sum (in terms of imaginary-time trajectories).
Furthermore, the improved computational efficiency provided by the minibatch method we introduced here allowed us to highlight the benefit of larger NNEs (i.e., longer trajectories) capable of accessing lower loss regions of state space using far less data than single NNs or small ensembles.

\section*{Code availability}
Our TPS implementation package is available through GitHub, \texttt{TransitionPathSampling.jl} \cite{MairTPS2022}, together with the source code to generate the figures and results in the paper \cite{MinibatchTPS2023}.

\begin{acknowledgments}
    We acknowledge 
    support from 
    EPSRC Grant no. EP/V031201/1
    and 
    University of Nottingham grant no.\ FiF1/3.   LC was supported by an EPSRC
    Doctoral prize from the University of Nottingham. 
    Simulations were performed using the University of Nottingham Augusta HPC cluster, and the Sulis Tier 2 HPC platform hosted by the Scientific Computing Research Technology Platform at the University of Warwick. Sulis is funded by EPSRC Grant EP/T022108/1 and the HPC Midlands+ consortium. We thank the creators and community of the Julia programming language \cite{julialang2017}, and acknowledge use of the packages \texttt{CUDA.jl} \cite{besard2018juliagpu,besard2019prototyping}, \texttt{Makie.jl} \cite{DanischKrumbiegel2021} and \texttt{ForwardDiff.jl} \cite{RevelsLubinPapamarkou2016}.
\end{acknowledgments}

\appendix
\section{Correction Distribution}
\label{app:correction-distribution}
Seita et al. \cite{Seita2018} show that we can calculate the $X_\text{corr}$ distribution numerically. We introduce a parameter $V$ to specify the range of values to sample the distribution over. We construct two discrete vectors $X$ and $Y$ with the elements of $X$ going linearly from $-2V$ to $+2V$ and $Y$ going from $-V$ to $+V$. The vector $X$ has $4N+1$ elements and the vector $Y$ has $2N+1$ elements.

From here, we define a matrix $M$ with elements
\begin{equation}
    M_{ij} = \Phi_\sigma(X_i - Y_j),
\end{equation}
where $\Phi_\sigma$ is the cumulative distribution function (CDF) of a normal distribution with variance $\sigma^2$. Additionally, we construct a new vector $v$ such that
\begin{equation}
    v_i = S(X_i),
\end{equation}
where $S$ is the logistic sigmoid function, i.e. the CDF of a logistically distributed random variable. Finally, we define the vector $u$ to be $u_j=C_\sigma(Y_j)$ which is our target to calculate. This can be calculated using the formula
\begin{equation}
    u = (M^T M + \lambda I)^{-1} M^T v,
\end{equation}
where $\lambda$ is regularisation parameter. We followed recommendations from Seita et al \cite{Seita2018} and used $V=10$, $N=4000$ and $\lambda=10$ to construct our numerical approximation of $C_\sigma$. We set any negative elements equal to zero and re-normalise the CDF to ensure the area under the curve will equal $1$.

Fortunately, the sampling algorithm allows us to calculate the distribution for a single value of $\sigma$ to save on computation and memory. This distribution can be calculated once and cached for future use. We can alter the acceptance condition to be
\begin{equation}
    \Delta^*(\trajparams, \trajparams') + X_{\text{nc}} + X_{\text{corr}} > 0,
\end{equation}
where $X_{\text{corr}}$ is sampled when $\sigma=1$ and $X_{\text{nc}}\sim \mathcal{N}(0, 1-\text{Var}[\Delta^*])$, requiring that $\text{Var}[\Delta^*]<1$.

\subsection*{Sampling the correction distribution}

The distribution can be efficiently sampled using the CDF: the cumulative sum of the probability distribution function (PDF). The CDF is a monotonically increasing set of $y$ values from $0$ to $1$. These values have corresponding $X$ values in the domain $-2V$ to $2V$. In order to sample this distribution we draw a random number, $u$, uniformly between $0$ and $1$. We find the $X$ which correponds to the intersection of $u=C_\sigma(X)$ by bisection on the discretised points and then linear interpolation between discretised values.
\section{Experiment Configurations}
\label{app:experiment-configuration}
Here, we provide the exact parameters used to generate data provided in the results.

\subsection{Linear Perceptron}
\label{app:linear-perceptron}
The linear perceptron model was trained on a simple $1$D problem, which was generated via $\bm{y} = m \bm{x}+ c$, where $x_i\sim \mathcal{U}(0, 1)$ and $m \sim \mathcal{U}(-1, 1)$ and $c \sim \mathcal{U}(-2, 0)$. We randomly sampled $256$ points to use in the distribution. The minibatch method used batch sizes of $32$ and set $\sigma = 0.1$ for the coupling between models in the trajectories. Experiments were run for $4\times 10^7$ epochs. Averages of observables were taken by discarding the first half of the data, to allow for a \textit{burn-in} time, and using minibatch estimates on the data.

\subsection{MNIST}
\label{app:mnist}

The convolutional neural network (CNN) model architecture used for our MNIST experiments was as follows:
\begin{enumerate}
    \item Input $28\times 28$ single channel image.
    \item Convolution layer with a $5 \times 5$ kernel and $16$ ouput channels.
    \item $2 \times 2$ max pooling layer.
    \item Convolution layer with a $3 \times 3$ kernel and $8$ ouput channels.
    \item $4 \times 4$ max pooling layer.
    \item Fully connected dense layer with $10$ outputs.
    \item Softmax layer to normalise probabilities.
\end{enumerate}
This model would output a normalised probability vector for each input image, specifying the ``liklihood'' of the image being a certain digit. The model contained $1906$, $32$-bit, floating point parameters.

For our experiments, we did not anneal the $s$ parameter, but instead, chose a fixed duration of $2\times 10^7$ epochs to allow for some ``burn-in'' time. The models were then run for another $2\times 10^7$ epochs, to allow for measuring the loss as an observable. Accuracies were only measured at the end of the $4\times 10^7$ epochs, due to the high computational demand.

We ran all of our experiments using $\sigma=0.05$ and only changed a random $25\%$ of the parameters of each model on each perturbation. Each perturbation changed only a single model in the trajectory, uniformly randomly.

We ran $6$ independent experiments for each set of presented parameters and calculated averages to present the results in \Cref{fig:mnist_combined_figure}, along with calculating the error bars using the averages' sample variance.
\section{Normality Investigation}
\label{app:normality-investigation}

\begin{figure}
    \centering
    \includegraphics[width=\columnwidth]{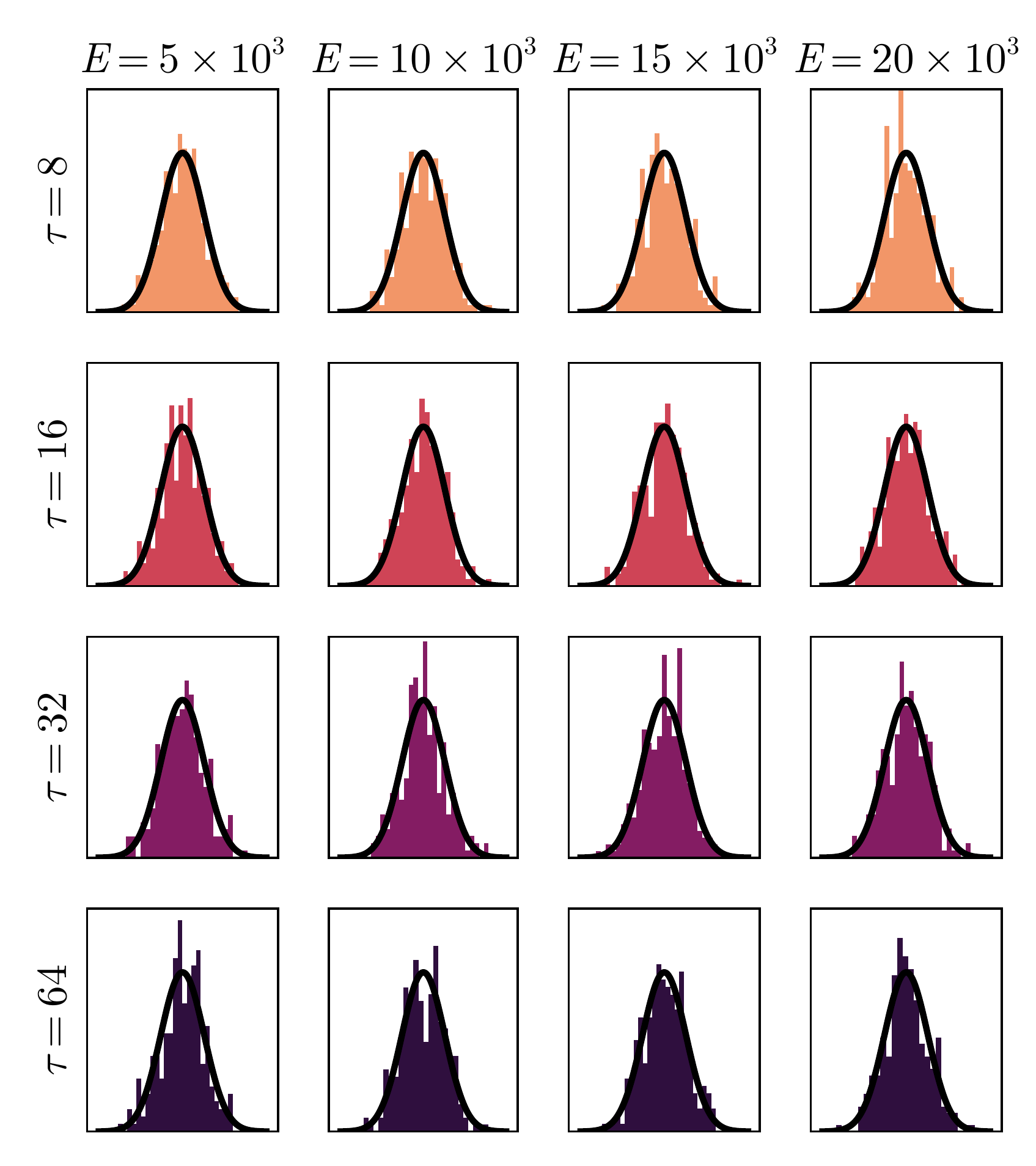}
    \caption{\label{fig:normality_investigation} A plot visually inspecting the normality of the $\Delta^*$ variable various values of $\tau$ at different specific epochs in the training process. This is done by sampling the delta loss for a single proposed change on the entire dataset of samples. These samples are split into minibatches of size $240$ and averages are taken, this gives a sample of $\Delta^*$. The distribution for the minibatches is plotted in each subfigure, along with a normal PDF (plotted as a solid black line) with a mean matching the true mean $\Delta$ and variance matching the empirical variance of the samples. Each row investigates a different $\tau$ and each column represents a different epoch in the training process, denoted by $E$.}
\end{figure}

To justify setting the error threshold $\delta \rightarrow \infty$, we ran a training experiment for two different $\tau$ at $s=50$ using a base batch size of $240$. These experiments were run for $20,000$ epochs and samples at intervals of $5,000$ epochs. Histograms of the individual $\Delta$ samples across the entire batch are plotted, along with a fitted curve showing the expected normal distribution given the empirical mean and variance of the batch data. This is presented in \Cref{fig:normality_investigation}.

\bibliography{references}

\end{document}